**The large-scale organization of metabolic networks**

H. Jeong[1], B. Tombor[2], R. Albert[1], Z. N. Oltvai[2] and A.-L. Barabási[1]

[1]Department of Physics, University of Notre Dame, Notre Dame, IN  46556, and

[2]Department of Pathology, Northwestern University Medical School, Chicago, IL  60611

**In a cell or microorganism the processes that generate mass, energy, information transfer, and cell fate specification are seamlessly integrated through a complex network of various cellular constituents and reactions [1]. However, despite the key role these networks play in sustaining various cellular functions, their large-scale structure is essentially unknown. Here we present the first systematic comparative mathematical analysis of the metabolic networks of 43 organisms representing all three domains of life. We show that, despite significant variances in their individual constituents and pathways, these metabolic networks display the same topologic scaling properties demonstrating striking similarities to the inherent organization of complex non-biological systems [2]. This suggests that the metabolic organization is not only identical for all living organisms, but complies with the design principles of robust and error-tolerant scale-free networks [2-5], and may represent a common blueprint for the large-scale organization of interactions among all cellular constituents.**

An important goal in biology is to uncover the fundamental design principles that provide the common underlying structure and function in all cells and microorganisms [6-13]. For example, it is increasingly appreciated that the robustness of various cellular processes is rooted in the dynamic interactions among its many constituents [14-16], such as proteins, DNA, RNA, and small molecules. Recent scientific developments improve our ability to identify the design principles that integrate these interactions into a complex system. Large-scale sequencing projects have not only provided complete sequence information for a number of genomes, but also allowed the development of integrated pathway-genome databases [17-19] that provide organism-specific connectivity maps of metabolic- and, to a lesser extent, various other cellular networks. Yet, due to the large number and the diversity of the constituents and reactions forming such networks, these maps are extremely complex, offering only limited insight into the organizational principles of these systems. Our ability to address in quantitative terms the structure of these cellular networks, however, has benefited from recent advances in understanding the generic properties of complex networks [2].

Until recently, complex networks have been modeled using the classical random network theory introduced by Erdös and Rényi (ER) [20, 21]. The ER model assumes that each pair of nodes (i.e., constituents) in the network is connected randomly with probability *p*, leading to a statistically homogeneous network, in which, despite the fundamental randomness of the model, most nodes have the same number of links, $\langle k \rangle$ (Fig. 1a). In particular, the connectivity follows a Poisson distribution strongly peaked at $\langle k \rangle$ (Fig. 1b), implying that the probability to find a highly connected node decays exponentially (i.e. $P(k) \sim e^{-k}$ for $k \gg \langle k \rangle$). On the other hand, recent empirical studies on the structure of the World-Wide Web [22], Internet [23], and social networks [2] have reported serious deviations from this random structure, demonstrating that these systems are described by scale-free networks [2] (Fig. 1c), for which $P(k)$ follows a power-law, i.e. $P(k) \sim k^{\gamma}$ (Fig. 1d). Unlike exponential networks, scale-free networks are extremely heterogeneous, their topology being dominated by a few highly connected nodes (hubs) which link the rest of the less connected nodes to the system (Fig. 1c). Since the distinction between the scale-free and exponential networks emerges as a result of simple dynamical principles [24, 25], understanding the large-scale structure of cellular networks can provide not only valuable and perhaps universal structural information, but could also lead to a better understanding of the dynamical processes that generated these networks. In this respect the emergence of power law distribution is intimately linked to the growth of the network in which new nodes are preferentially attached to already established nodes [2], a property that is also thought to characterize the evolution of biological systems [1].

To begin to address the large-scale structural organization of cellular networks, we have examined the topologic properties of the core metabolic network of 43 different organisms based on data deposited in the WIT database [19]. This integrated pathway-genome database predicts the existence of a given metabolic pathway based on the annotated genome of an organism combined with firmly established data from the biochemical literature. As 18 of the 43 organisms deposited in the database are not yet fully sequenced, and a substantial portion of the identified ORFs are functionally unassigned, the list of enzymes, and consequently the list of substrates and reactions (see Table 1 in Supplementary Material [26]), will certainly be expanded in the future. Nevertheless, this publicly available database represents our current best approximation for the metabolic pathways in 43 organisms and provides sufficient data for their unambiguous statistical analysis (see Methods and Supplementary Material [26]).

As we illustrate in Fig. 1e, we have first established a graph theoretic representation of the biochemical reactions taking place in a given metabolic network. In this representation, a metabolic network is built up of nodes, which are the substrates that are connected to one another through links, which are the actual metabolic reactions. The physical entity of the link is the temporary educt-educt complex itself, in which enzymes provide the catalytic scaffolds for the reactions yielding products, which in turn can become educts for subsequent reactions. This representation allows us to systematically investigate and quantify the topologic properties of various metabolic networks using the tools of graph theory and statistical mechanics [21]. Our first goal was to identify the structure of the metabolic networks, i.e., to establish if their topology is best described by the inherently random and uniform exponential model [21] (Fig. 1a and b), or the highly heterogeneous scale-free model [2] (Fig. 1c and d). As illustrated in Fig. 2, our results convincingly indicate that the probability that a given substrate participates in *k* reactions follows a power-law distribution, i.e., metabolic networks belong to the class of scale-free networks. Since under physiological conditions a large number of biochemical reactions (links) in a metabolic network are preferentially catalyzed in one direction (i.e., the links are directed), for each node we distinguish between incoming and outgoing links (Fig. 1e). For instance, in *E. coli* the probability that a substrate participates as an educt in $k$ metabolic reactions follows $P(k) \sim k^{-\gamma_{in}}$, with $\gamma_{in}$ = 2.2, and the probability that a given substrate is produced by $k$ different metabolic reactions follows a similar distribution, with $\gamma_{out}$ = 2.2 (Fig. 2b). We find that scale-free networks describe the metabolic networks in all organisms in all three domains of life (Fig. 2a-c) [26], indicating the generic nature of this structural organization (Fig. 2d).

A general feature of many complex networks is their small-world character [27], i.e., any two nodes in the system can be connected by relatively short paths along existing links. In metabolic networks these paths correspond to the biochemical pathway connecting two substrates (Fig. 3a). The degree of interconnectivity of a metabolic network can be characterized by the network diameter, defined as the shortest biochemical pathway averaged over all pairs of substrates. For all non-biological networks examined to date the average connectivity of a node is fixed, which implies that the diameter of a network increases logarithmically with the addition of new nodes [2, 27, 28]. For metabolic networks this implies that a more complex bacterium with higher number of enzymes and substrates, such as *E. coli*, would have a larger diameter than a simpler bacterium, such as *M. genitalium*. In contrast, we find that the diameter of the metabolic network is the same for all 43 organisms, irrespective of the number of substrates found in the given species (Fig. 3b). This is surprising and unprecedented, and is possible only if with increasing organism complexity individual substrates are increasingly connected in order to maintain a relatively constant metabolic network diameter. Indeed, we find that the average number of reactions in which a certain substrate participates increases with the number of substrates found within the given organism (Fig. 3c and d).

An important consequence of the power-law connectivity distribution is that a few hubs dominate the overall connectivity of the network (Fig. 1c), and upon the sequential removal of the most-connected nodes the diameter of the network rises sharply, the network eventually disintegrating into isolated clusters that are no longer functional. Yet, scale-free networks also demonstrate unexpected robustness against random errors [5]. To examine if metabolic networks display a similar error tolerance we performed computer simulations on the metabolic network of the bacterium, *E. coli*. Upon removal of the most connected substrates the diameter increases rapidly, illustrating the special role these metabolites play in maintaining a constant metabolic network diameter (Fig. 3e). However, when randomly chosen *M* substrates were removed, -mimicking the consequence of random mutations of catalyzing enzymes-, the average distance between the remaining nodes was not affected, indicating a striking insensitivity to random errors. Indeed, *in-silico* and *in-vivo* mutagenesis studies indicate a remarkable fault tolerance upon removal of a substantial number of metabolic enzymes from the *E. coli* metabolic network [29]. Of note, data similar to that shown in Fig. 3e have been obtained for all investigated organisms, without detectable correlations with their evolutionary position.

Since the large-scale architecture of the metabolic network rests on the most highly connected substrates, we need to address whether the same substrates act as hubs in all organisms, or if there are major organism-specific differences in the identity of the most connected substrates. When we rank order all the substrates in a given organism based on the number of links they have (Table 1)[26], we find that the ranking of the most connected substrates is practically identical for all 43 organisms. Also, only ~4% of all substrates that are found in all 43 organisms are present in all species. These substrates represent the most highly connected substrates found in any individual organism, indicating the generic utilization of the same substrates by each species. In contrast, species-specific differences among various organisms emerge for less connected substrates. To quantify this observation, we examined the standard deviation ($\sigma_r$) of the rank for substrates that are present in all 43 organisms. As shown in Fig. 3f, we find that $\sigma_r$ increases with the average rank order, $\langle r \rangle$, implying that the most connected substrates have a relatively fixed position in the rank order, but the ranking of less connected substrates is increasingly species-specific. Thus, the large-scale structure of the metabolic network is identical for all 43 species, being dominated by the same highly connected substrates, while less connected substrates preferentially serve as educt or product of species-specific enzymatic activities.

The contemporary topology of a metabolic network reflects a long evolutionary process molded in general for a robust response towards internal defects and environmental fluctuations and in particular to the ecological niche the specific organism occupies. As a result, one expects that these networks are far from being random, and our data demonstrate that the large-scale structural organization of metabolic networks is indeed highly similar to that of robust and error-tolerant networks [2, 5]. The uniform network topology observed in all 43 organisms strongly suggests that, irrespective of their individual building blocks or species-specific reaction pathways, the large-scale structure of metabolic networks is identical in all living organisms, in which the same highly connected substrates may provide the connections between modules responsible for distinct metabolic functions [1].

A unique feature of metabolic networks, as opposed to that seen in non-biological scale-free networks, is the apparent conservation of the network diameter in all living organisms. Within the special characteristics of living systems this attribute may represent an additional survival and growth advantage, since a larger diameter would attenuate the organism's ability to efficiently respond to external changes or internal errors. For example, should the concentration of a substrate suddenly diminish due to mutation in its main catalyzing enzyme, offsetting the changes would involve the activation of longer alternative biochemical pathways, and consequently the synthesis of more new enzymes, than within a smaller metabolic network diameter.

But how generic these principles are for other cellular networks (e.g., information transfer, cell cycle)? While the current mathematical tools do not allow unambiguous statistical analysis of the topology of other networks due to their relatively small size, our preliminary analysis suggest that connectivity distribution of non-metabolic pathways also follows a power-law distribution, indicating that cellular networks as a whole are scale-free networks. Therefore, the evolutionary selection of a robust and error tolerant architecture may characterize all cellular networks, for which the scale-free topology with a conserved network diameter appears to provide an optimal structural organization.

**Methods**

*Database preparation:* For our analyses of core cellular metabolisms we used the "Intermediate metabolism and Bioenergetics" portions of the WIT database [19] (http://igweb.integratedgenomics.com/IGwit/), that predicts the existence of a metabolic pathway in an organism based on its annotated genome (i.e., on the presence of the presumed open reading frame (ORF) of an enzyme that catalyzes a given metabolic reaction), in combination with firmly established data from the biochemical literature. As of December 1999, this database provides description for 6 archaea, 32 bacteria and 5 eukaryota. The downloaded data were manually rechecked, removing synonyms and substrates without defined chemical identity.

*Construction of metabolic network matrices:* Biochemical reactions described within a WIT database are composed of substrates and enzymes connected by directed links. For each reaction, educts and products were considered as nodes connected to the temporary educt-educt complexes and associated enzymes. Bi-directional reactions were considered separately. For a given organism with $N$ substrates, $E$ enzymes and $R$ intermediate complexes the full stochiometric interactions were compiled into an $(N+E+R) \times (N+E+R)$ matrix, generated separately for each of the 43 organisms.

*Connectivity distribution* [$P(k)$]: Substrates generated by a biochemical reaction are products, and are characterized by incoming links pointing to them. For each substrate we have determined $k_{in}$, and prepared a histogram for each organism, providing how many substrates have exactly $k_{in}$ =0,1,…. Dividing each point of the histogram with the total number of substrates in the organism provided $P(k_{in})$, or the probability that a substrate has $k_{in}$ incoming links. Substrates that participate as educts in a reaction have outgoing links. We have performed the analysis described above for $k_{in}$, determining the number of outgoing links ($k_{out}$) for each substrate. To reduce noise logarithmic binning was applied.

*Biochemical pathway lengths* [$\Pi(l)$]: For all pairs of substrates, the shortest biochemical pathway, $\Pi(l)$ (i.e., the smallest number of reactions by which one can reach substrate B from substrate A) were determined using a burning algorithm. From $\Pi(l)$ we determined the diameter, $D = \sum_{l} l \cdot \Pi(l) / \sum_{l} \Pi(l)$, which represents the average path length between any two substrates.

*Substrate ranking* [ $<r>_o$, $\sigma(r)$ ]: Substrates present in all 43 organisms (i.e., a total of 51 substrates) were ranked based on the number of links each had in each organisms, having considered incoming and outgoing links separately (*r* =1 were assigned for the substrate with the largest number of connections, and *r* =2 for second most connected one, etc.). This way for each substrate a well-defined *r* value in each organism was defined. The average rank $<r>_o$ for each substrate was determined by averaging *r* over the 43 organisms. We also determined the standard deviation, $\sigma(r) = <r^2>_o - <r>_o^2$ for all 51 substrates present in all organisms.

*Analysis of the effect of database errors:* Of the 43 organisms whose metabolic network we have analyzed the genome of 25 has been completely sequenced (5 Archaea, 18 Bacteria, 2 Eukaryotes), while the remaining 18 are only partially sequenced. Therefore two major sources of possible errors in the database could affect our analysis: (a) the erroneous annotation of enzymes and consequently, biochemical reactions; for the organisms with completely sequenced genomes this is the likely source of error. (b) reactions and pathways missing from the database; for organisms with incompletely sequenced genomes both (a) and (b) are of potential source of error. We investigated the effect of database errors on the validity of our findings, the results being presented in the Supplementary Material [26], indicating that the results offered in this paper are robust to these errors.

**Acknowledgement**

We would like to acknowledge all members of the WIT project for making this invaluable database publicly available for the scientific community. We also thank C. Waltenbaugh and H. S. Seifert for comments on the manuscript. Research at the University of Notre Dame was supported by the National Science Foundation, and at Northwestern University by grants from the National Cancer Institute.

Correspondence and requests for materials should be addressed to A.-L.B. (alb@nd.edu) or Z.N.O. (zno008@northwestern.edu).

**FIGURE LEGENDS**

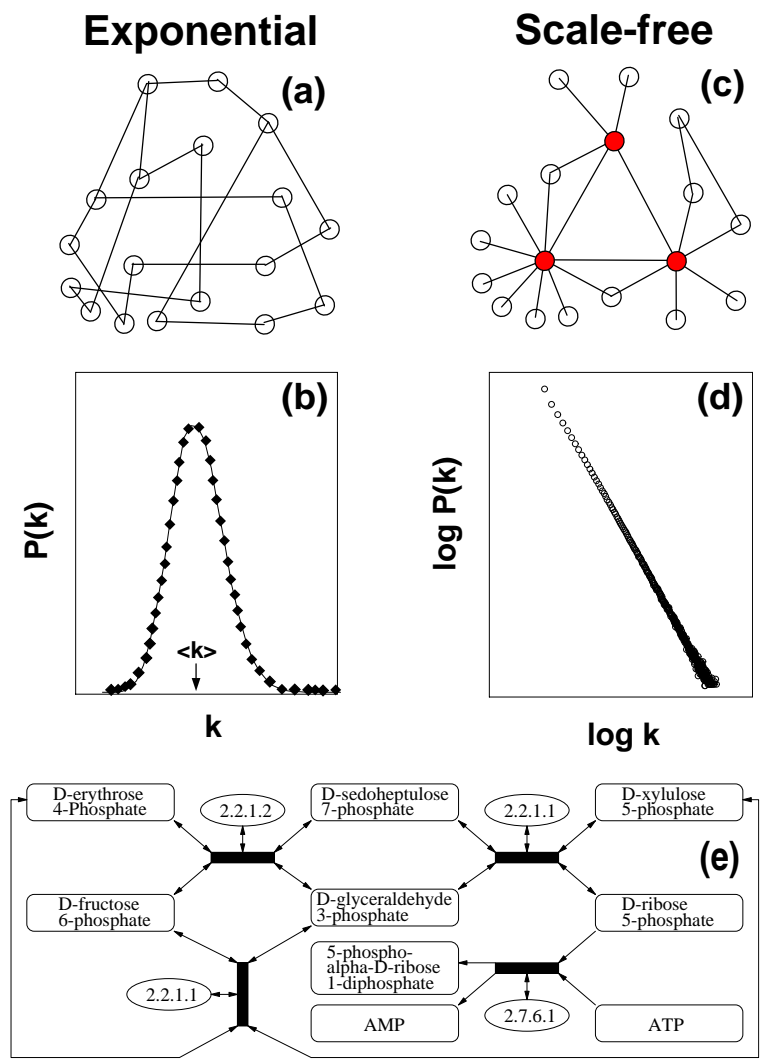

**Figure 1**

(**a**) Representative structure of the network generated by the classical random network model of Erdös and Rényi. (**b**) The network connectivity can be characterized by probability, *P(k),* that a node has *k* links. For a random network *P(k)* is strongly peaked at $k = \langle k \rangle$ and decays exponentially for large *k* (i.e. $P(k) \sim e^{-k}$ for $k \gg \langle k \rangle$ and $k \ll \langle k \rangle$). (**c**) In the scale-free network most nodes have only a few links, but a few nodes, called hubs (red), have a very large number of links. (**d**) *P(k)* for a scale-free network has no well-defined peak, and for large *k*, it decays as a power-law, $P(k) \sim k^{-\gamma}$, appearing as a straight line with slope -γ on a log-log plot. (**e**) A portion of the WIT DB for the bacterium, *E. coli*. Each substrate can be represented as a node of the graph, linked to one another through temporary educt-educt complexes (black boxes) from which the products emerge as new nodes (substrates). The enzymes which provide the catalytic scaffolds for the reactions, are shown by their EC numbers.

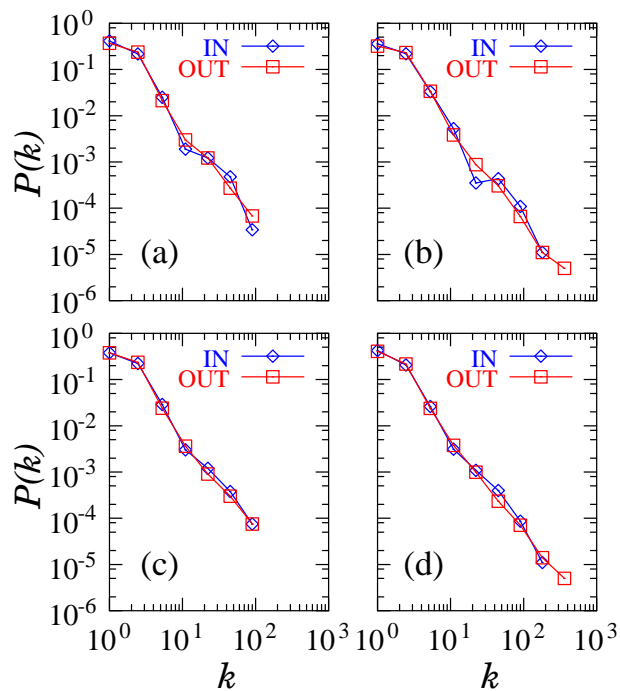

**Figure 2**

Connectivity distribution *P(k)* for the substrates in (**a**) *A. fulgidus* (Archae) (**b**) *E. coli* (Bacterium) (**c**) *C. elegans* (Eukaryote), shown on a log-log plot, counting separately the incoming (IN) and outgoing links (OUT) for each substrate, $k_{in}$ ($k_{out}$) corresponding to the number of reactions in which a substrate participates as a product (educt). The characteristics of the three organisms shown in **a-c** and the exponents $\gamma_{in}$ and $\gamma_{out}$ for all organisms are given in Table 1 [26]. (**d**) The connectivity distribution averaged over all 43 organisms.

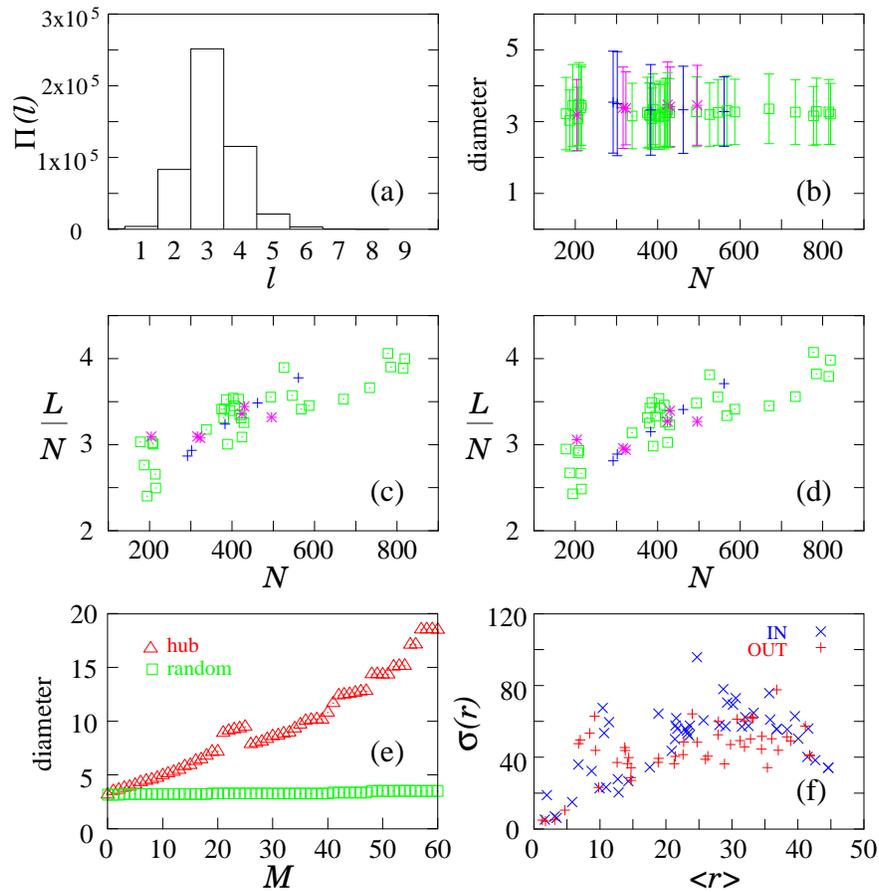

**Figure 3**

(**a**) The histogram of the biochemical pathway lengths, $\ell$, in the bacterium, *E. coli*. (**b**) The average path length (diameter) for each of the 43 investigated organisms. The error bars correspond to the standard deviation $\sigma \sim \langle\ell^2\rangle - \langle\ell\rangle^2$ as determined from $\Pi(\ell)$ (shown in (a) for *E. coli*). (**c**) The average number of incoming links or (**d**) outgoing links per node for each organism. (**e**) The effect of substrate removal on the metabolic network diameter of the bacterium, *E. coli*. In the upper curve ($\triangle$) in an inverse order of connectivity, the most connected substrates are removed first. In the bottom curve ($\square$) nodes are removed randomly. *M*=60 corresponds to ~8% of the total number of substrates in found in *E. coli*. (**f**) Standard deviation of the substrate ranking ($\sigma_r$) as a function of the average ranking, $\langle r \rangle_o$ for substrates present in all 43 investigated organisms. The horizontal axis in (b,c,d,) denotes the number of nodes in each organism. Archaea (magenta), bacteria (green), and eukaryotes (blue) are shown in (b,c,d,f).